\documentclass[]{aa}  

\usepackage{graphicx}
\usepackage{txfonts}
\usepackage[utf8]{inputenc}
\usepackage{graphics}
\usepackage{amssymb}
\usepackage{lscape}
\usepackage{epsfig}
\usepackage{natbib}
\usepackage{longtable}
\usepackage{hyperref}
\usepackage[T1]{fontenc}
\usepackage[utf8]{inputenc}
\usepackage{multicol}
\usepackage[dvipsnames]{xcolor}

\usepackage{lipsum}

\usepackage[normalem]{ulem}

\usepackage{caption}
\usepackage{subcaption}

\usepackage{booktabs}
\usepackage{hyperref}
\usepackage{multirow}
\usepackage{enumitem}

\usepackage[switch]{lineno}

\usepackage[dvipsnames]{xcolor}
\definecolor{myblue}{HTML}{1F77B4}
\definecolor{mygreen}{HTML}{2CA02C}
\definecolor{m}{HTML}{D62728}
\definecolor{mymagenta}{HTML}{D33682}
\definecolor{codepurple}{HTML}{C42043}

\newcommand{\Msol}{\(M_\odot\)}

\hypersetup{
unicode=true,           
colorlinks=true,        
citecolor=myblue,       
urlcolor=black        
}
\begin{document} 

   \title{AGNs in massive galaxy clusters: Role of galaxy merging, infalling groups, cluster mass, and dynamical state}


   \author{E. Koulouridis\inst{1}
          \and
          A. Gkini\inst{2,3}
          \and
          E. Drigga\inst{1,4}
          }

   \institute{Institute for Astronomy \& Astrophysics, Space Applications \& Remote Sensing, National Observatory of Athens, GR-15236 Palaia Penteli, Greece\\
              \email{ekoulouridis@noa.gr}
         \and
             Department of Astronomy, The Oskar Klein Centre, Stockholm University, AlbaNova, SE-106 91 Stockholm, Sweden
         \and
         Department of Astrophysics, Astronomy \& Mechanics, Faculty of Physics, National and Kapodistrian University of Athens, Panepistimiopolis Zografou, Athens 15784, Greece
         \and
         Sector of Astrophysics, Astronomy \& Mechanics, Department of Physics, Aristotle University of Thessaloniki, Thessaloniki 54124, Greece
             }
\authorrunning{E. Koulouridis et al.}
\titlerunning{AGNs in massive galaxy clusters}
   \date{\today}

 
  \abstract
   {There is compelling evidence that active galactic nuclei (AGNs) in high-density regions have undergone a different evolution than their counterparts in the field, indicating that they are strongly affected by their environment. However, we still lack a comprehensive understanding of the dominant mechanisms that trigger the nucleus and the processes that drive the evolution of AGNs in clusters.}
   {To investigate (and possibly disentangle) the various factors that may affect the prevalence of AGNs in cluster galaxies, we selected a sample of 19 thoroughly studied X-ray-selected galaxy clusters from the LoCuSS survey. All these clusters are considered massive, with $M_{500}\gtrsim 2\times10^{14} M_\sun$, and span a narrow redshift range between $z\sim0.16$ and 0.28.}
   {We divided the cluster surroundings into two concentric annuli with a width of $R_{500}$ radius. We considered the first annulus as the central cluster region and the second as the outskirts. We further divided the cluster sample based on the presence of infalling X-ray-detected groups, cluster mass, or dynamical state. We determined the AGN fraction in cluster galaxies of the various sub-samples by correlating the X-ray point-like sources selected from the 4XMM DR10 catalogue with the highly complete spectroscopic catalogue of cluster members obtained with Hectospec. We subsequently used the optical spectra to determine the type of nuclear activity and we visually inspected the host morphology for indications of galaxy mergers or other interactions.}
   {We found that the X-ray AGN fraction in the outskirts is consistent with the field, but it is significantly lower in cluster centres, in agreement with previous results for massive clusters. We show that these results do not depend on cluster mass, at least within our cluster mass range, nor on the presence of X-ray-detected infalling groups. Furthermore, we did not find any evidence of a spatial correlation between infalling groups and AGNs. Nevertheless, a significant excess of X-ray AGNs is found in the outskirts of relaxed clusters at the 2$\sigma$ confidence level, compared both to non-relaxed clusters and to the field. Finally, according to the literature, the fraction of broad- to narrow-line AGNs in clusters is roughly consistent with the field. However, broad-line AGNs may be preferably located in cluster centres. In the outskirts, the optical spectra of X-ray AGNs present narrow emission lines or they are dominated by stellar emission.}
   {Our results suggest that the mechanisms that trigger AGN activity may vary between cluster centres and the outskirts. Ram pressure can efficiently remove the gas from infalling galaxies, thereby triggering AGN activity in some cases. However, the reduced availability of gas globally diminishes the fraction of AGNs in cluster centers. The surplus of X-ray AGNs identified in the outskirts of relaxed clusters may be attributed to an increased frequency of galaxy mergers, a notion that is further supported by the disturbed morphology observed in several galaxies.}


   \maketitle
%
\section{Introduction}

As a consequence of hierarchical structure formation, the majority of galaxies eventually fall into clusters \citep[e.g.][]{Eke04,Calvi11}. Therefore, clusters are the principal environment of galaxies and they can play a very important role in galaxy evolution. In addition, supermassive black holes (SMBHs)  also appear to be key elements in galaxy evolution, as much in the local \citep[e.g.][]{Magorrian1998,Ferrarese2000,Kormendy2013} as well as in the distant Universe \citep{Yang2019}. They are essentially found in all massive galaxies and they are easier to study when they manifest as immensely emitting active galactic nuclei (AGNs). However, deciphering the behaviour of AGNs and studying their demographics is not a trivial task. They are not only complicated objects in terms of their phenomenology, but also intrinsically variable and very compact, making them unsuitable for direct observation. Furthermore, there is strong evidence that AGNs are affected by their environment, both locally at the level of their host galaxy and its immediate surroundings \citep{Maiolino1997,Koulouridis2006b,Koulouridis2013,Dultzin2008,Manzer2014,Silva2021,Duplancic2021,Pierce2023,Li2023} as well as at the level of large-scale structures, from voids \citep{Constantin2008,Mishra2021,Ceccarelli2022} to galaxy clusters \citep[e.g.][]{Koulouridis2010,Stroe2020,Munoz2023}, and superclusters \citep{Koulouridis16b}. Therefore, it is crucial to thoroughly investigate the AGN population of galaxy clusters, as both the immense structure and the powerful nucleus seem to play an important role in galaxy evolution, however, the interplay among these key elements is not well understood.   

This uncertainty is due to the various physical mechanisms that may affect galaxies and SMBHs within clusters. Several studies have shown that massive clusters ($M>10^{14}M_\sun$) can effectively suppress the fraction of AGNs in cluster galaxies \citep[e.g.][]{Kauffmann2004,Gavazzi2011,Ehlert2013,Ehlert2014,Mishra2020,Beyoro2021}, probably through ram pressure stripping (RPS). This can lead to a deficit of the cold gas reservoir available to trigger the nuclear activity \citep[e.g.][]{Gunn72,Cowie77,Giovanelli85,Popesso06b,Chung2009,Haines2012,Sabater2013,Jaffe15,Poggianti2017b} and its effect is expected to be proportional to the mass of the cluster and inversely proportional to the mass of the affected galaxy \citep[e.g.][]{Boselli2022}. Indeed, RPS is probably not so effective in poor clusters and groups, where AGN activity in galaxies is at least as frequent as in the field \citep{Sabater2012, Koulouridis14, Koulouridis2018b}. However, \citet{Poggianti2017b} suggested that RPS may also act as a triggering mechanism for AGN activity in cluster members. More recently, \citet{Peluso2022} confirmed that the so-called "jellyfish" galaxies \citep{Chung2009,Bekki2009,Poggianti2017b} host a significantly higher number of AGNs than similar galaxies in the field.

On the other hand, an increase in the AGN activity in the outskirts of clusters has been reported in several studies \citep[e.g.][]{Johnson2003,Branchesi2007,Koulouridis14}, although the results vary depending on the different selection of clusters and AGN samples. In particular, \citet{Ruderman2005} discovered a mild excess of X-ray sources between 1.5 and 3 Mpc in massive clusters spanning the redshift range of z$ = 0.3 - 0.7$. However, the excess was found only in dynamically relaxed clusters, while no excess was found in the outskirts of disturbed clusters. These findings were confirmed more recently in the optical band by \citet{Stroe2021}, in a sample of 14 clusters ($z\sim 0.15-0.31$), spanning a wide range of masses and dynamical states. They found that the H$\alpha$-detected AGN fraction peaks in the outskirts of relaxed clusters ($\sim 1.5-3$ Mpc). In addition, \citet{Koulouridis2018b} found a significant overdensity of spectroscopically confirmed X-ray AGNs in the outskirts of low-mass clusters (M$_{500,MT} < 10^{14} M_\sun$ and $0.1<z< 0.5$) from the XXL Survey \citep{Adami2018}, while no excess was confirmed for higher cluster masses. At higher redshifts, a similar excess of X-ray AGNs was also reported by \citet{Fassbender2012} between 4 and 6 arcmin from the centres of 22 massive clusters ($0.9 < z < 1.6$). \citet{Koulouridis2019} confirmed a highly significant excess  of X-ray point-like sources in the outskirts ($2-2.5R_{500}$) of the five most massive, $M_{500}^{SZ}>10^{14} M_{\odot}$ , and distant, z$\sim$1, galaxy clusters in the \textit{Planck} and South Pole Telescope (SPT)\textit{} surveys. Very recently, an AGN excess in the outskirts ($\sim3R_{500}$) was also found in the {\it Magneticum} simulations \citep{Rihtarsic2023}. Such an increase may be interpreted by a corresponding increase in galaxy merging rate, which is favoured by the lower galaxy velocities in the outskirts when compared to the centre, as well as low-mass groups when compared to massive clusters \citep[e.g.][]{Ehlert15, Lopes2017, Gordon2018}. Another possibility is that AGNs enter the cluster environment along with infalling small groups that offer a more favourable environment for AGN triggering (pre-processed). On the contrary, some studies have  found that AGNs have no special position inside galaxy clusters \citep[e.g.][]{Gilmour2009, Ehlert15}, unless we only consider the most powerful optical AGNs. Furthermore, \citet{Bufanda2017} found no differences in the fraction of luminous X-ray AGNs ($L_{\rm X}>10^{43}$ erg sec$^{-1}$) between groups and clusters for 432 clusters from the Dark Energy Survey (DES) up to $z=0.95$. 

Finally, many studies have revealed a positive evolution of the AGN fraction in cluster galaxies with redshift. Above a redshift of $z\sim $1, the AGN fraction seems to surpass the corresponding field density \citep[e.g.][]{Fassbender12,Kocevski2009,Martini2013,Bufanda2017,Hashiguchi2023}. In addition, low-mass protoclusters at higher redshifts may potentially contain a higher number of AGNs \citep[][]{Lehmer13,Krishnan17}

Considering the various mechanisms reported above and the degeneracies among them, it is evident that a more detailed and comprehensive analysis is needed based on various well defined cluster samples. In this context, we investigated the presence of X-ray-selected AGNs in a sample of 19 massive X-ray-detected clusters within a narrow redshift bin. Massive galaxy clusters represent the deepest gravitational potentials of the Universe and, as such, they are unique laboratories for studying the effect of dense environment on AGN triggering and evolution. The novelty of this study lies on the wealth of available multi-wavelength data for these 19 clusters that have enabled accurate estimations of cluster and AGN properties. Furthermore, the X-ray selection presents some important advantages both for AGNs and clusters. First, it has been well established that the most efficient way to detect AGNs is through X-ray observations \citep[e.g.][]{Brandt15}. Second, X-ray-detected cluster samples trace the hot gas trapped in deep gravitational potentials and, thus, they are much less affected by projection effects than optically selected samples \citep[e.g.][]{Ramos-Ceja2019}. Most importantly, the gas density and temperature structure of the inter-cluster medium (ICM) can be used to derive cluster mass profiles under the assumption that the ICM is in hydrostatic equilibrium \citep[e.g.][]{Frenk1990}. \citet{Koulouridis2019} argued that an accurate estimate of cluster mass (and, therefore, of the associated cluster physical size) can be very important when stacking AGN number counts from multiple clusters and from various regions where physical conditions and processes may differ.   

The outline of this paper is as follows. In Sect. \ref{sec:sample} we discuss the data preparation and sample selection. The methodology is described in Sect. \ref{sec:method}.  In Sect. \ref{sec:results} we present our results.  Our discussion and conclusions are presented in Sects. \ref{sec:disc} and \ref{sec:conc}, respectively.  Throughout this paper, we assume a Planck cosmology \citet{Planck2016} of $h=0.678$, H$_{o}=100$ h km s$^{-1}$ Mpc$^{-1}$, $\Omega_{m}=0.308,$ and $\Omega_{\Lambda}=0.691$. The cluster masses are hydrostatic. 

\section{Sample selection}
\label{sec:sample}

\begin{figure*}
\centering
        \includegraphics[width=9cm]{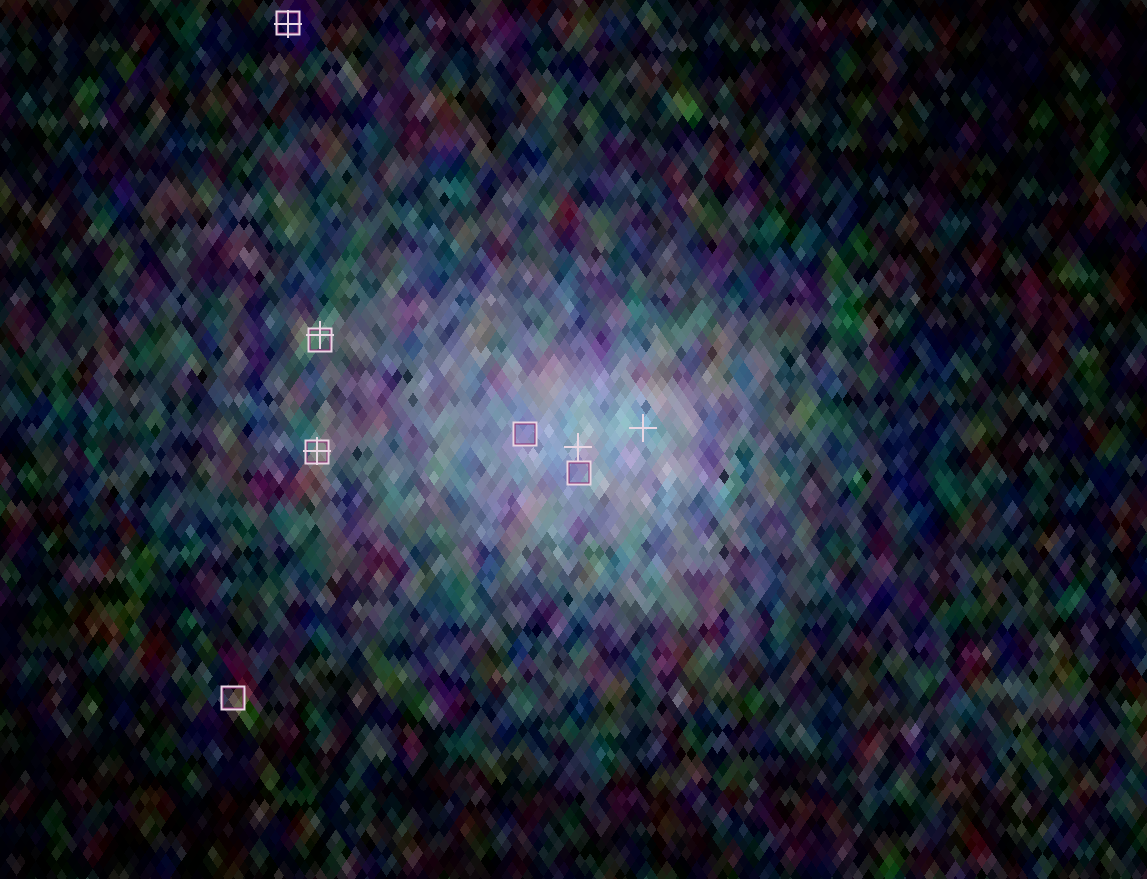}
        \includegraphics[width=9cm]{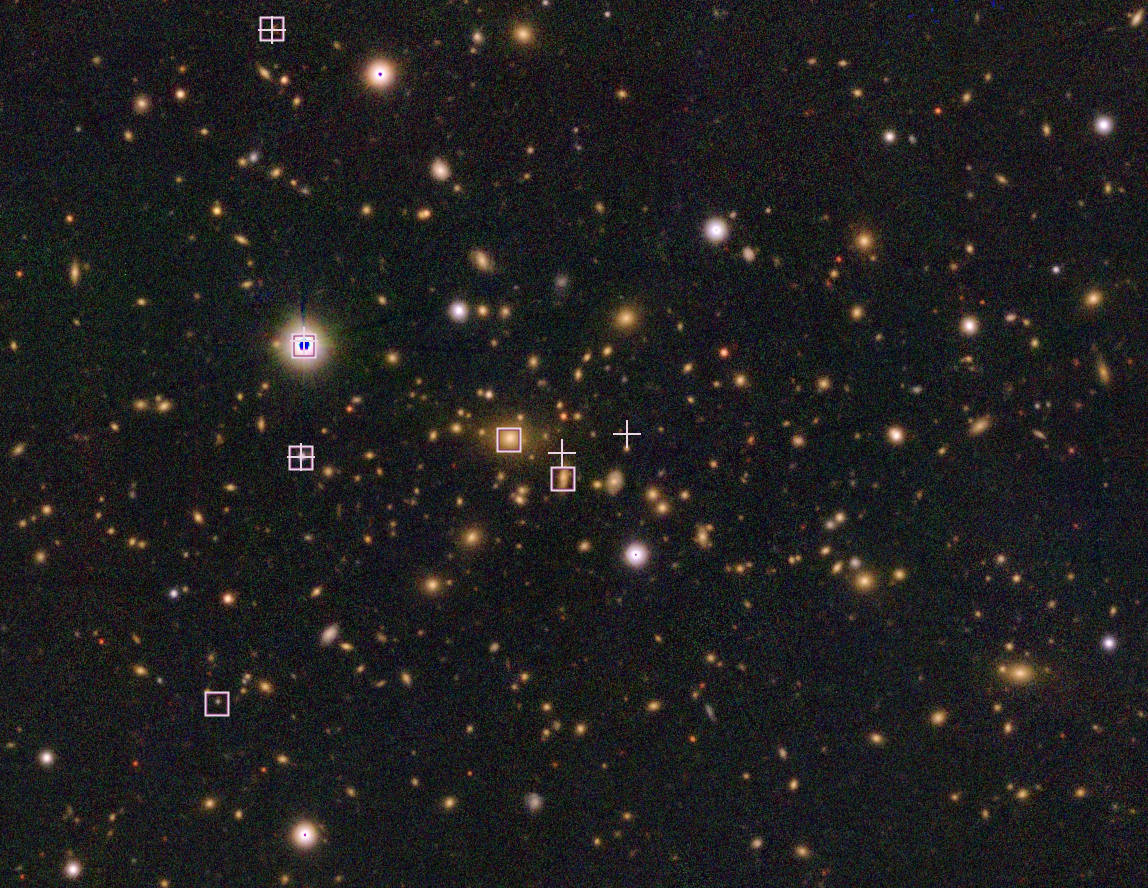}
        \caption{X-ray ({\it XMM-Newton}, left panel) and  optical (Pan-STARRS, right panel) images of the central cluster region within 0.5$R_{500}$ of Abell 1763. Crosses and squares mark {\it XMM-Newton} and {\it Chandra} detections, respectively. Sources with a flux lower than the threshold applied in sect.~\ref{sect:detections} are ommited. {\it Chandra} detects a point-like X-ray source in close proximity to the brightest cluster galaxy (BCG) and at concordant redshift with the cluster. All other sources detected by either telescope or both are discarded due to their redshift or extent, as described in Sect.~\ref{sect:detections}.}
        \label{fig:Chandra}
\end{figure*} 

\begin{table}[t]
  \centering
  \caption{List of clusters and corresponding {\it XMM-Newton} observations.\\(1) Cluster name. (2) {\it XMM-Newton} observation ID. (3) Duration in seconds}
  \label{observ}
  \begin{tabular}{lcc}
  \hline
  \noalign{\smallskip}
  Cluster &  XMM obs. ID & Duration \\
  name & & (sec) \\
  (1)&(2)\\
  \hline
  \noalign{\smallskip}
        \vspace{4pt}
    Abell 68 & 0084230201 & 30108 \\
            \vspace{4pt}
    Abell 209 & 0084230301 &24709 \\
    Abell 611 & 0781590301 & 22800\\ 
    &0781590501 & 21700\\ 
    &0781590201 & 23000\\ 
    &0781590401 & 21700\\
           \vspace{4pt}
 &0605000601 &36285\\
    Abell 697 &  0605000701 &30809\\ 
           \vspace{4pt}
 & 0827041001 &36400\\
         \vspace{4pt}

    Abell 963 & 0084230701 &27461\\
            \vspace{4pt}
    Abell 1758 & 0142860201 &57217\\
            \vspace{4pt}
    Abell 1763 & 0084230901 &26937\\
    Abell 1835 & 0551830201 & 120870\\
    &0098010101&61353\\
    &0147330201&106844\\
                      \vspace{4pt}
   &0551830101&121671\\
           \vspace{4pt}

    Abell 1914 & 0112230201&25815\\
            \vspace{4pt}
    ZwCl1454.8+2233 & 0108670201 & 46705\\
    Abell 2219 & 0112231801 & 17720  \\
    &0112231901&17471\\
                \vspace{4pt}
    &0605000501&19915\\
    RXJ1720.1+2638 & 0500670201& 30409\\
    &0500670301&24415\\
    &0500670401&23372\\
            \vspace{4pt}
    Abell 2390 & 0111270101 & 23105 \\
            \vspace{4pt}
    ZwCl0104.4+0048&0762870601&29500\\
            \vspace{4pt}
    Abell 383&0084230501&33645\\
    Abell 586&0673850201&19918\\
    &0605000801&20327\\
           \vspace{4pt}
    &0827051101&29000\\
    ZwCl0857.9+2107&0402250401&13269\\
           \vspace{4pt}
    &0402250701&16192\\
    Abell 1689&0093030101&39763\\
        \vspace{4pt}
    RXJ2129.6+0005&0093030201&58916\\
    \hline
    
  \end{tabular}
\end{table}   

\begin{figure}
\centering
        \includegraphics[width=9.5cm]{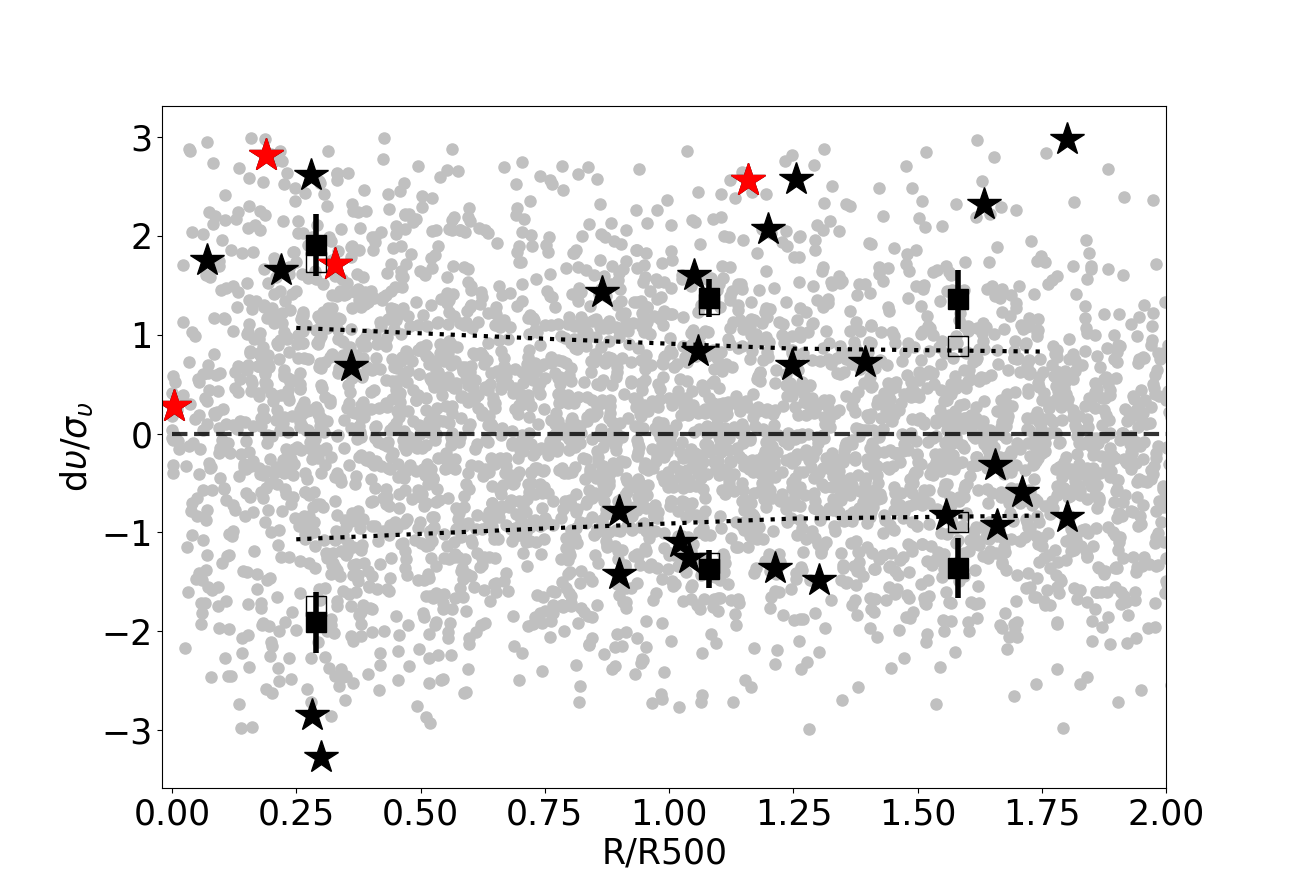}
        \caption{$d\upsilon/\sigma_\upsilon$ versus $R/R_{500}$ diagram. Grey circles mark cluster galaxies, while the dotted line indicates their velocity dispersion. Stars mark the position of X-ray-detected AGN and the red colour indicates those that exhibit broad emission lines in their optical spectra. Filled squares denote the AGN velocity dispersion in three bins of equally divided number of sources. Open squares mark the respective median values. Close to the cluster cores, X-ray-detected AGNs tend to have higher velocities with respect to the cluster than the rest of the population. This trend may indicate the presense of infalling galaxies prior to their first pericenter passage. The average AGN velocity exhibits a declining trend toward the outskirts, mirroring that of cluster galaxies.}
        \label{fig:du}
\end{figure}

\begin{table*}[t]
  \centering
  \caption{Properties of clusters and X-ray-detected infalling groups.\\(1) Cluster name, (2) X-ray-detected group name as in \citet{Haines2018}, (3) right ascension, (4) declination, (5) average cluster (boldface) or group redshift, (6) cool-core cluster designation based on \citep{Bianconi2021}, (7)-(8) cluster $R_{500}$ and hydrostatic $M_{500}$ as computed in \citet{Martino2014}, (9)-(10) computed group $R_{200}$ from Equation \ref{eq:1} using $M_{200}$ from \cite{Haines2018}. }
  \label{clusters}
  \begin{tabular}{lcccccccccc}
  \hline
  \noalign{\smallskip}
  Cluster & Group & RA & Dec & z  & CC & $R_{500}$ & $M_{500}$ & $R_{200}$ & $M_{200}$\\
  name & name & (J2000)  & (J2000)  &  & & (Mpc) & (10$^{14}$ \Msol) & (Mpc) & (10$^{14}$ \Msol) \\
  (1)&(2)&(3)&(4)&(5)&(6)&(7)&(8)&(9)&(10)\\
  \hline
  \noalign{\smallskip}
    Abell 68 & & 00:37:06.84 & +09:09:24.28 & \bf 0.251 & N & 1.40 & 10.44 &  \\
      & A68-g7 & 00:37:38.77 & +09:05:10.5 & 0.245 &&&& 0.638 & 0.357   \\
    Abell 209 & & 01:31:53.45 & -13:36:47.84 & \bf0.209 & N & 1.15 & 5.45 & \\
    & A209-g3 & 01:31:35.32 & -13:31:54.4 & 0.207 &&&& 0.758 & 0.572 \\
    & A209-g6 & 01:32:05.17 & -13:39:53.4 & 0.212 &&&& 0.827 & 0.747 \\
    & A209-g10 & 01:31:37.04 & -13:45:01.9 & 0.200 &&&& 0.615 & 0.303 \\
    Abell 611 && 08:00:56.81 & +36:03:23.40 & \bf0.286 & N &1.20 & 6.80  &  \\
    & A611-g5 & 08:01:25.69 & +36:05:37.0 & 0.284 &&&& 0.819 & 0.788 \\
    & A611-g6 & 08:01:11.04 & +36:05:21.8 & 0.289 &&&& 0.910 & 0.109\\
    Abell 697 && 08:42:57.58 & +36:21:59.54 & \bf0.282 & N &1.50 & 13.14  &   \\
    & A697-g4 & 08:43:00.79 & +36:28:41.3& 0.280 &&&& 0.813 & 0.765 \\
    Abell 963 && 10:17:03.65 & +39:02:49.63 & \bf0.204 & Y & 1.14 & 5.60  &  \\
    & A963-g1 & 10:17:08.24 & +38:50:58.1 & 0.199 &&&& 0.629 & 0.325 \\
   & A963-g5 & 10:16:35.43 & +39:10:05:7 & 0.209 &&&& 0.556  & 0.227 \\
    & A963-g10 & 10:16:40.13 & +38:54:43.0 & 0.201  &&&& 0.915 & 1.00  \\
    Abell 1758 && 13:32:44.66 & +50:30:26.49 &\bf 0.279 & N & 1.38 & 10.21  &  \\
    & A1758-g7 & 13:32:10.67 & +50:30:31.5 & 0.279  &&&& 0.658  & 0.406 \\
     & A1758-g8 & 13:32:31.43 & +50:24:37.1 & 0.272 &&&& 1.761 & 7.724 \\
    Abell 1763 && 13:35:18.07 & +40:59:57.16 &\bf 0.232 & N & 1.33 & 6.60  &   \\
     & A1763-g5 & 13:35:35.04 & +41:06:02.4 & 0.236 &&&& 0.734 & 0.536 \\
     & A1763-g6 & 13:36:12.27 & +41:04:55.3 & 0.236  &&&& 0.799 & 0.694  \\
     & A1763-g7 & 13:34:52.77 & +40:57:02.6 & 0.237 &&&& 1.162 & 2.130\\
     & A1763-g9 & 13:35:05.44 & +40:58:29.7 & 0.237 &&&& 0.956 & 1.189 \\
     & A1763-g11 & 13:35:33.39 & +40:52:08.1 & 0.229 &&&& 0.671 & 0.407 \\
    Abell 1835 && 14:00:52.50 & +02:52:42.64 & \bf0.252 & Y & 1.57 & 14.04 &  \\
     & A1835-g9 & 14:00:31.66 & +02:54:39.1 & 0.250 &&&& 0.685 & 0.443 \\
     & A1835-g11 & 14:00:21.25 & +02:50:18.0 & 0.251 &&&& 0.580 & 0.269 \\
    & A1835-g13 & 14:01:23.72 & +02:46:43.6 & 0.246 &&&& 0.602 & 0.299  \\
    & A1835-g14 & 14:01:16.15 & +02:45:13.2 & 0.245 &&&& 0.763 & 0.608\\
    Abell 1914 && 14:25:59.70 & +37:49:41.63 &\bf 0.167 & N & 1.38 & 8.08 &  \\
    & A1914-g6 & 14:25:16.54 & +37:45:01.6 & 0.170 &&&& 0.669 & 0.379 \\
     & A1914-g7 & 14:25:03.97 & +37:57:30.0 & 0.162 &&&& 0.912 & 0.95 \\
    ZwCl1454.8+2233 && 14:57:15.11 & +22:20:34.26 & \bf0.257 & Y & 1.06 & 3.65  &  \\
     & Z7160-g5 & 14:57:41.14  & +22:23:33.2 & 0.253 &&&& 0.623 & 0.334  \\ 
    Abell 2219 && 16:40:22.56 & +46:42:21.60 & \bf0.226 & N & 1.75 & 14.35  &    \\
    & A2219-g5 & 16:40:10.88 & +46:48:51.4 & 0.235 &&&& 0.729 & 0.526 \\
     & A2219-g7 & 16:40:07.17 & +46:36:29.5 & 0.222 &&&& 0.747 & 0.557\\
    RXJ1720.1+2638 && 17:20:10.14 & +26:37:30.90 &\bf 0.160 & Y & 1.23 & 6.97  &  \\
     &R1720-g4 &17:19:26.09 &+26:33:33.0 & 0.154 &&&& 0.522 & 0.177 \\
     &R1720-g6 & 17:20:02.23 &+26:29:44.8 & 0.161 &&&& 0.724 & 0.474 \\
    Abell 2390 && 21:53:36.85 & +17:41:43.66 & \bf0.229 & Y & 1.59 & 13.67  &   \\
    & A2390-g1 & 21:53:09.47 & +17:42:24.9 & 0.222 &&&& 0.695 & 0.449 \\
    ZwCl0104.4+0048 & & 01:06:49.50 & +01:03:22.10 &  \bf0.253 & Y  &  0.76 &  1.67  \\
    Abell 383 &&  02:48:03.42 & -03:31:45.05 & \bf 0.189 & Y &  1.01 &   3.25   \\
    Abell 586 &&  07:32:20.22 & +31:37:55.88 &\bf 0.171 & N & 1.08 &  4.42   \\
    ZwCl0857.9+2107 &&  09:00:36.86 & +20:53:39.84 & \bf0.234 & Y  & 0.91 & 2.33   \\
    Abell 1689 &&  13:11:29.45 & -01:20:28.32 &\bf 0.185 & N &  1.51 & 11.98  \\
    RXJ2129.6+0005 & & 21:29:39.88 & +00:05:20.54 & \bf0.234 & Y&  1.08 & 4.22   \\
    \hline
  \end{tabular}
\end{table*}

\subsection{X-ray-detected galaxy clusters}

In the current study we used a sample of 19 X-ray-selected massive clusters, of $M_{500}\gtrsim 2\times10^{14} M_\sun$, from the Local Cluster Substructure Survey (LoCuSS), which span a narrow redshift range between $z\sim0.16$ and 0.28. The sample considered here is an unbiased subsample of the complete high-L$_{X}$ LoCuSS sample with available high quality wide-field multi-wavelength data. For more information about the LoCuSS sample, we refer to \citet{Bianconi2021} and references therein. The 19 clusters are listed in Table \ref{observ}, along with the respective {\it XMM-Newton} observations. Most importantly, these 19 clusters have been thoroughly analysed in an earlier series of papers, which allows us to  simultaneously investigate various aspects of this multi-faceted problem. In more detail, \citet{Martino2014} applied a very detailed fitting of the available X-ray data and derived the hydrostatic mass of each cluster, within three different radii, $R_{\Delta}$,
with $\Delta$ = 2500, 1000, and 500, where $R_{\Delta}$ is the clustercentric radius containing the mass of $M_{\Delta} = \Delta\rho 4/3 \pi R_{\Delta}^3$ , with $\rho$ as the critical density of the Universe at the redshift of the cluster. For our purposes, we used the $R_{500}$ radius for a direct comparison of our results with those of previous studies. In addition, \citet{Bianconi2021} applied a similarly detailed X-ray fit on the {\it Chandra} data and classified these clusters into cool core (CC) and non-cool core (non-CC) classes. Furthermore, \citet{Haines2018} analysed the {\it XMM-Newton} X-ray observations covering the area around these clusters and reported all the X-ray-detected galaxy groups. These groups were spectroscopically confirmed at the redshift of the central cluster with a mass as low as 2\% of the central massive cluster and were considered as infalling structures. The galaxy groups are also listed in Table \ref{clusters} under their corresponding cluster. The R$_{200}$ of the groups were calculated using the following expression:

\begin{equation}
\label{eq:1}
    R_{200} = [\frac{3 M_{200}}{4 \pi}\frac{1}{200 \rho_{c,0} [\Omega_{m,0}(1+z)^{3}+\Omega_{\Lambda}]}]^{1/3},
\end{equation}
where $\rho_{c,0}$ is the critical density of the universe today, $\Omega_{m,0}$ is the density parameter of the matter today, and $\Omega_{\Lambda}$ is the density parameter of the dark energy.

\subsection{X-ray-detected AGNs}\label{sect:detections}

To select the X-ray point-like sources around each massive cluster, we used the 4XMM Data Release 10 (DR10) catalogue \citep{Webb2020}. All 19 clusters were targeted observations by {\it XMM-Newton}, which means that they were laying at the centre of the detectors. For each cluster, we first selected all the X-ray sources up to a maximum distance of 13 arcmin to avoid the external region of the {\it XMM} detectors where the vignetting effect is very prominent and occasionally, there is only partial PN detector coverage. To ensure that we can reach sources down to a uniform luminosity lower limit in all 19 observations, we computed the sensitivity limit of all observations using XMM Science Analysis System (SAS) v20.0.0. All clusters have deep X-ray observations to reach as low as a luminosity threshold of $L_{X[0.5-10]\,\rm{keV}} > 10^{42}$ erg s$^{-1}$ at the redshift of each cluster. This luminosity threshold signifies that the X-ray emission from point-like sources is most probably from an AGN -- and not, for example, from X-ray binaries or star formation. All X-ray sources below this limit were discarded.

From the above analysis, we excluded the central cluster region where the X-ray background is very high because of the diffuse cluster emission, which may hinder the detection of point-like sources. This area is approximately contained within 0.5$R_{500}$ radii in all instances, constituting 25\% of the total area of the central annulus. However, we employed the corresponding {\it Chandra} X-ray observations for this region, which are not significantly impacted by this effect. By utilizing {\it Chandra} observations, we identified and included six missed AGNs in the central cluster regions and confirmed those previously detected by {\it XMM}. Any {\it XMM} point-source detection within 0.5$r_{500}$ radii lacking confirmation from corresponding {\it Chandra} detection was excluded. Furthermore, any sources exhibiting extended X-ray emission ($>10''$) were excluded, as these likely represent clumps of thermal gas. The chosen threshold resulted from a visual inspection of optical images for all sources with reported non-zero extent in the 4XMM catalog. An illustrative example is provided in Fig.~\ref{fig:Chandra}, where a narrow-line AGN detected exclusively by Chandra was included in Abell 1763.  

\section{Methodology}
\label{sec:method}

The large majority (90\%) of cluster galaxies below a magnitude limit of $M_K^*+1.5$ were observed with the Hectospec spectrograph mounted on the 6.5m MMT telescope in the framework of the Arizona Cluster Redshift Survey (ACReS)\footnote{\url{http://herschel.as.arizona.edu/acres/acres.html}}. The highly complete spectroscopy ensures that the vast majority of X-ray-detected AGNs that indeed belong to the cluster have an optical counterpart with a spectroscopically confirmed redshift. All X-ray sources were cross-correlated with cluster galaxies of the ACReS project and the hosts were selected within 4 arcsec from the X-ray position. This radius was selected considering the {\it XMM} point-spread-function (PSF) and visual inspection of all candidate optical counterparts. We used {\it Aladin} sky atlas \citep{Boch2014} in order to visualize images of the hosts from PanSTARRS\footnote{\url{https://outerspace.stsci.edu/display/PANSTARRS/}}, DES\footnote{\url{https://www.darkenergysurvey.org/}}, or DECaLS\footnote{\url{https://www.legacysurvey.org/decamls/}} depending on the coverage of each cluster location. In some cases, also imaging by the Hyper Suprime Camera (HSC)\footnote{\url{https://hsc.mtk.nao.ac.jp/ssp/}} of the Subaru Telescope was available, while some AGNs located in cluster centres have also HST imaging.  All clusters were found within the redshift range of $0.16<z<0.29.$  Therefore, in most cases, the cluster members are shown to have a very good imaging resolution.

Host galaxies were included as cluster members if their relative line-of-sight (LoS) velocity offset, $d\upsilon/\sigma_\upsilon$ was less than three, where $d\upsilon$ is the relative velocity of the galaxy with respect to the cluster, and $\sigma_\upsilon$ is the cluster velocity dispersion. X-ray AGNs were found to have relatively higher velocities than cluster member galaxies \citep{Haines2012}, but they usually do not surpass this threshold. Velocity dispersions used in the current study were taken from the literature \citep{Haines2015} based on cluster galaxies within $R_{200}$. X-ray point-like sources that were cross-correlated with galaxies outside this limit or not cross-correlated with a cluster member (thus considered to be hosted in fainter galaxies than the ACReS threshold) were discarded as projections. In Table~\ref{AGNlist} we present the final list of 30 X-ray-detected AGNs and in Fig.~\ref{fig:du}, we plot their locations on the $d\upsilon/\sigma_\upsilon$ vs. $R/R_{500}$ plane, along with non X-ray-detected cluster galaxies. It is important to acknowledge that in the central annulus, the detection of low-luminosity sources may be impeded within the diffuse X-ray emission, resulting in a reduced count of AGNs despite our efforts to mitigate this effect using {\it Chandra} observations. On the other hand, in the same region, approximately 30\% of the X-ray sources and 15\% of member galaxies could be projected from outside the $R_{500}$ radius, resulting in an increase of $\sim$20\% in the computed AGN fraction. The uncertainty introduced by these two counter-acting effects is considerably smaller than the confidence limits for small number of events in astrophysical data \citep{Gehrels86} applied throughout the current work.

To study the effect of cluster environment on AGN activity, we divided the area into concentric annuli centred on the X-ray peak of the cluster emission. Each annulus has a width of $R_{500}$ radius and depending also on redshift the XMM field of view (13 arcmin considered in the current study) covers a radius that varies in each cluster from 1.5 to 3$R_{500}$. The $R_{500}$ values were estimated in \citet{Martino2014} by fitting the profile of the cluster X-ray emission. The use of any $R_\Delta$ radius is critical when investigating the effect of cluster environment on AGNs and their host galaxies, since it can be directly linked to the physical conditions at the location of each galaxy. Most importantly, we can assume that roughly the same conditions exist within the specific annulus of other clusters independent of the actual physical or projected distance. The importance of a well measured $R_{500}$ radius was demonstrated in \citet{Koulouridis2019}.

The cluster sample in the current paper was divided based on: (i) cluster mass, (ii) the existence of infalling X-ray-detected groups,  and (iii) the presence of a cool core as a proxy of their dynamical state. For more details on the estimation of cluster mass and entropy (to define CC and non-CC clusters), we can refer to \citet{Bianconi2021} and \citet{Sanderson2009}. We then count the number of cluster members in each annulus that host X-ray AGNs, as defined in the previous section. To obtain significant results, AGN members found in the same annulus in all clusters of each subsample were first added and then divided with the total number of spectroscopically confirmed cluster galaxies found in the same annuli by the ACReS survey. The result is the AGN fraction in cluster galaxies in each annulus. To make a comparison with the respective AGN fraction in field galaxies, we used the relevant value reported in \citet{Haggard2010} for sources with X-ray luminosity $L_x>10^{42}$ hosted in field galaxies with absolute magnitudes of $M_R<-20$ between $z=0.05$ and $z=0.31$. This magnitude limit is consistent with the galaxy selection in the current paper, as they both reach $M^*+1.5$ \citep[see][]{Haines2012}. Finally, we also inspected the optical spectra of all AGNs to investigate any correlation between spectral class and cluster properties.

\section{Results}
\label{sec:results}

\subsection{Effect of cluster mass} \label{with}

\begin{figure}
\centering
        \includegraphics[scale=0.6]{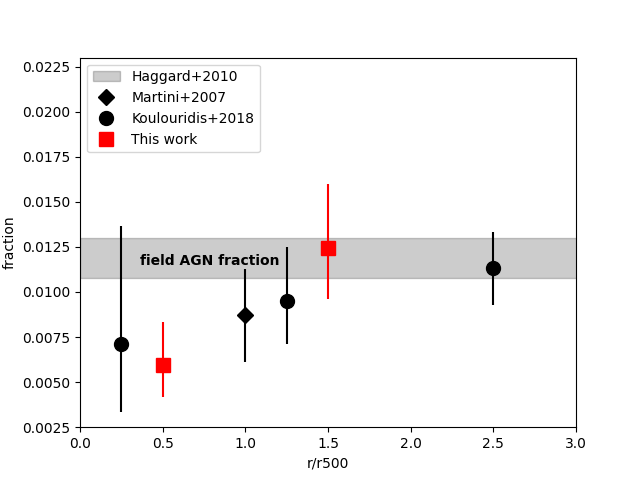}
        \caption{Fraction of cluster galaxies ($M_K<M_K^*+1.5$) hosting an X-ray AGN ($L_{\rm X [0.5-10\,keV]}>10^{42}$ erg sec$^{-1}$). The results are plotted as a function of distance from the cluster centre. Error bars indicate the 1$\sigma$ confidence limits for small numbers of events \citep{Gehrels86}. For comparison, we plot the results from previous analyses on clusters of similar mass and redshift.}
        \label{fig:all}
\end{figure}

The sample used in the current analysis comprises clusters within the mass range of $ 2\times10^{14}\;M_\sun <M_{500}< 2\times10^{15}\;M_\sun$. First, we present the results of the full sample and we compare them with the literature. The fraction of X-ray AGNs with $L_x>10^{42}$ erg s$^{-1}$ in cluster galaxies is presented in Fig.~\ref{fig:all}. We found that the AGN fraction within $R_{500}$ is $\sim$50\% lower, at the $2\sigma$ confidence level, than in field galaxies as estimated in \citet{Haggard2010}. This is in agreement with the results of previous studies on similarly massive clusters \citep[e.g.][]{Koulouridis2010, Martini07,Haines2012}. However, in the second annulus, which corresponds to 1-2 $R_{500}$ radius, the AGN fraction is already consistent with the field, again in excellent agreement with a similar analysis of a large sample of X-ray-detected clusters \citep{Koulouridis2018b} in the the XXL survey \citep{Pierre2016,Adami2018}. 

\begin{figure}
\centering
        \includegraphics[scale=0.6]{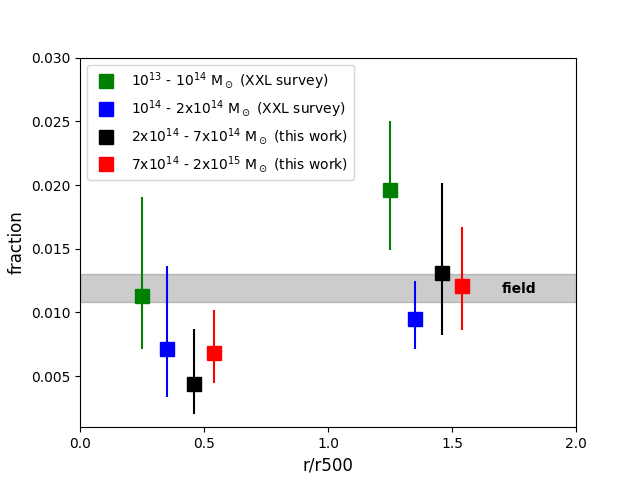}
        \caption{Fraction of cluster galaxies ($M_K<M_K^*+1.5$) hosting an X-ray AGN ($L_{\rm X [0.5-10\,keV]}>10^{42}$ erg sec$^{-1}$). The results are plotted as a function of distance from the cluster centre. The sample is divided in two based on cluster mass. Error bars indicate the 1$\sigma$ confidence limits for small numbers of events \citep{Gehrels86}. For comparison, we plot results from the analysis of massive clusters by \citet[][namely:\ 167 clusters, 0.1$<z<$0.5]{Koulouridis2018b}. A significant AGN excess is found at the 95\% confidence level only in the outskirts of low mass clusters ($M<10^{14}M_\sun$).}
        \label{fig:mass}
\end{figure} 

\begin{figure}
\centering
        \includegraphics[scale=0.55]{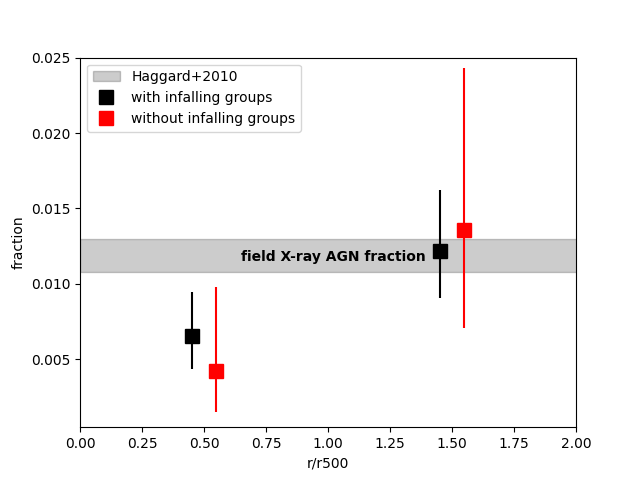}
        \caption{Fraction of cluster galaxies ($M_K<M_K^*+1.5$) hosting an X-ray AGN ($L_{\rm X [0.5-10\,keV]}>10^{42}$ erg sec$^{-1}$). The results are plotted as a function of distance from the cluster centre. The sample
is divided based on the presence of X-ray-detected infalling groups. Error bars indicate the 1$\sigma$ confidence limits for small numbers of events \citep{Gehrels86}.}
        \label{fig:infall}
\end{figure} 

\begin{figure*}
\centering
        \includegraphics[scale=0.5]{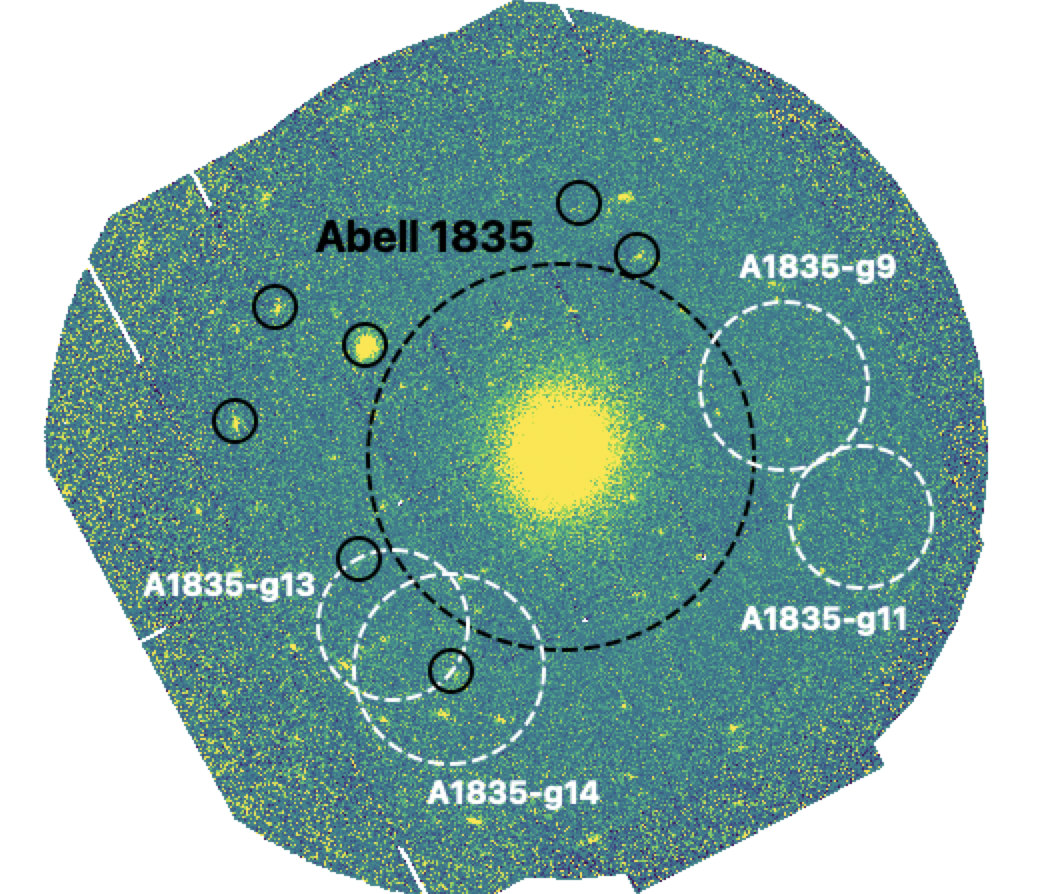}\includegraphics[scale=0.5]{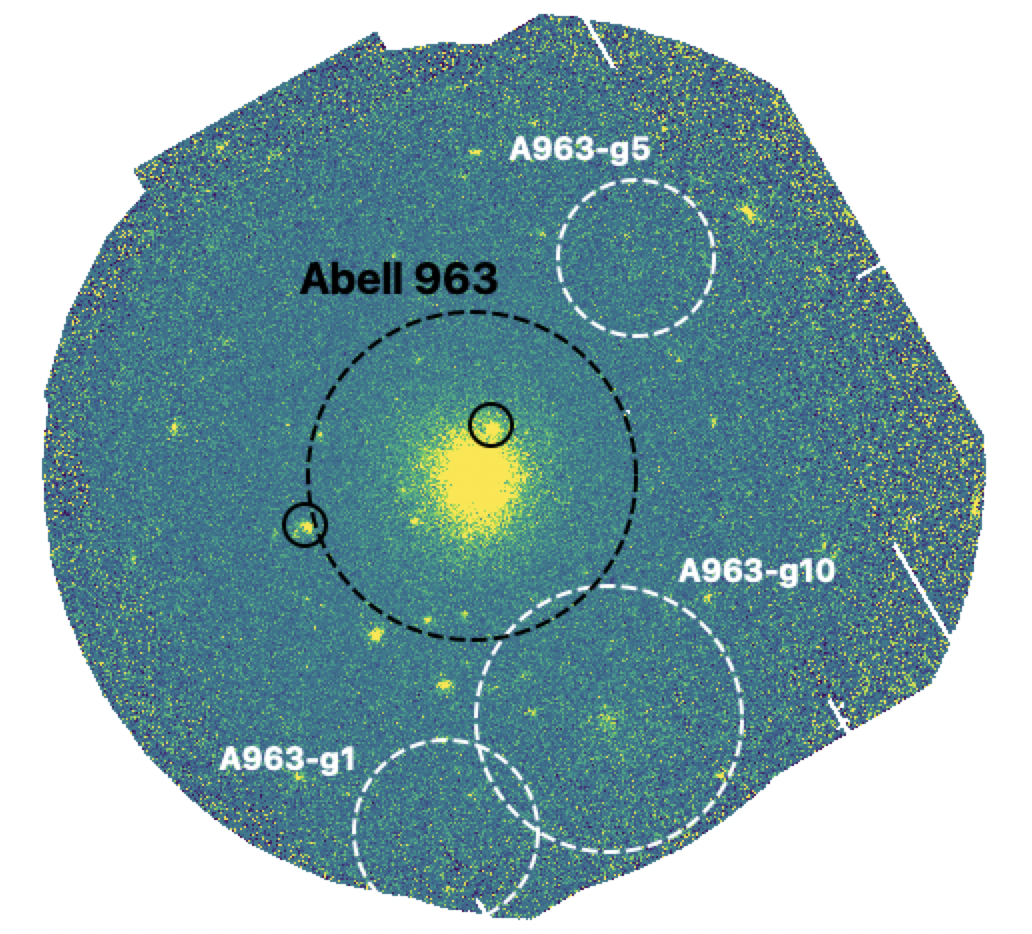}
        \caption{{\it XMM-Newton} observations in the [0.5 -- 2]\,keV X-ray band of Abell 1835 (left panel) and Abell 963 (right panel).  The images are not background-subtracted and their diameter is $15'$. The confirmed X-ray-detected AGNs are marked by black solid circles and the R$_{500}$ of the cluster by black dashed circle. The white dashed circles of diameter R$_{200}$ indicate the groups that are infalling into the clusters.}
        \label{fig:clusters}
\end{figure*} 

 We further divided the sample according to cluster mass using the $M_{500}$ hydrostatic mass measurements by \citet{Martino2014} who combined data from both {\it Chandra} and {\it XMM} observations. To divide the sample in two equal parts, we selected the value of $M_{500}=7\times 10^{15} M_\sun$, which corresponds to $M_{200}=10^{15} M_\sun$ by extrapolation of the mass profiles out to $R_{200}$ \citep[see][]{Haines2018}. We note, however, that even the less massive half of the sample contains clusters with a mass $M_{500}>2\times 10^{14} M_\sun$, which places them above the regime of groups and poor clusters. The main results are presented in Fig.~\ref{fig:mass}. We also plot the results from \citet{Koulouridis2018b} from a similar analysis of XXL survey clusters. The mass range of the XXL sample is complementary to the one used here, covering the range of $10^{13}\;M_\sun <M_{500}< 2\times10^{14}\;M_\sun$. It is evident that the fraction of AGNs in clusters with $M_{500}>10^{14} M_\sun$ is similar in all samples. However, there is significant excess of X-ray AGNs in the outskirts of poor clusters and groups ($M<10^{14}M_\sun$), while in group centres, the AGN fraction is still consistent with the field, in contrast with the well established suppression observed in more massive clusters.

\subsection{Effect of infalling groups}

To investigate the role of possible recent infall of galaxy groups in the potential of massive clusters, we used the results of \citep{Bianconi2021} to divide the sample into clusters with and without X-ray-detected galaxy groups. All groups are spectroscopically confirmed. The results are presented in Fig.~\ref{fig:infall}. The subsample that has no X-ray-detected infalling groups comprises of only six clusters that are mostly those with the lowest masses. Thus, the number of AGNs detected in this subsample is low and the uncertainties are large. Nevertheless, there is no evidence that the fraction of X-ray AGNs is affected by the presence of infalling groups. Both subsamples have an AGN fraction that is roughly consistent with the field in the outskirts and they also present the same decrease towards the centre.  

In addition, we investigated the position of X-ray AGNs relatively to the detected groups. Examples are shown in Fig. \ref{fig:clusters}. In general, very few AGNs are found close to the position of groups. Visual inspection of the optical images showed that AGNs are not located in local galaxy overdensities that may indicate the presence of small groups not emitting in X-rays. Nevertheless, X-ray AGNs do indeed comprise an infalling population in galaxy clusters, as made evident from their relative velocities with respect to the clusters \citep[see][]{Haines2012}. However, they are not related to groups, as we  also confirmed with a two-sample Kolmogorov-Smirnov (KS) test. In particular, we compiled simulated catalogues of galaxy groups in each XMM observation with random positions. Each sample comprises ten times more random groups than the actual detected ones. Then we computed the distances between groups and X-ray AGNs, both for the random and the real X-ray-detected groups, and we applied a KS test. The result indicates that we cannot reject the null hypothesis that the samples originate from the same parent distribution with any statistical significance.  

\subsection{Effect of the cluster dynamical state}

Following the analysis in \citet{Bianconi2021} we also divided our sample into clusters with either a low- or high-entropy core. Entropy is a proxy of the dynamical state of the cluster \citep{Voit2005, Sanderson2009, Ghirardini2017}. A low-entropy core signifies a cool-core cluster (CC), while a non-cool-core cluster (non-CC) exhibits a significantly higher entropy \citep[e.g.][]{Sanderson2009}. There is a clear dichotomy between CC and non-CC clusters, with the two categories demonstrating significantly different, not only entropy but also temperature and and gas density profiles.

\begin{figure}
\centering
        \includegraphics[scale=0.55]{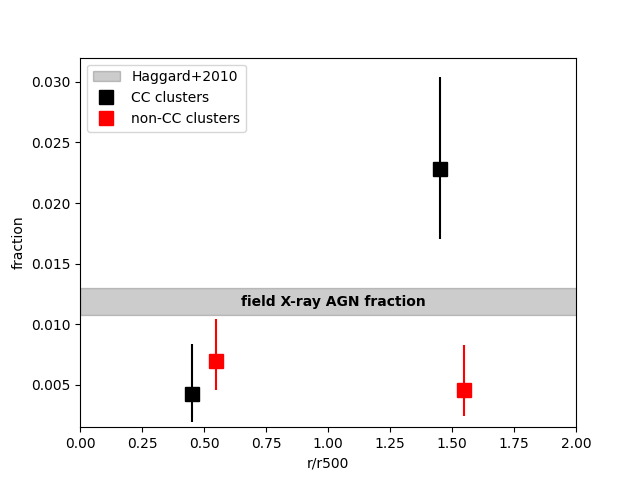}
        \caption{Fraction of cluster galaxies ($M_K<M_K^*+1.5$) hosting an X-ray AGN ($L_{\rm X [0.5-10\,keV]}>10^{42}$ erg sec$^{-1}$). The results are plotted as a function of distance from the cluster centre. The sample
is divided in cool-core (CC) and non-cool-core (non-CC). Error bars indicate the 1$\sigma$ confidence limits for small numbers of events \citep{Gehrels86}.}
        \label{fig:entropy}
\end{figure}

Our cluster sample is divided almost equally to two sub-samples of low and high entropy clusters. For more information on CC and non-CC clusters, we refer to more specialized articles \citep[e.g.][]{Vikhlinin2006,Sanderson2009,McDonald2019}. 
The AGN fractions of CC and non-CC clusters are presented in Fig.~\ref{fig:entropy}. In this case, a large difference is evident in the outskirts of clusters, with CC clusters presenting a significantly higher fraction of X-ray AGNs than non-CC clusters and field galaxies at the 2$\sigma$ confidence level.     

\begin{figure}
\centering
        \includegraphics[width=8cm]{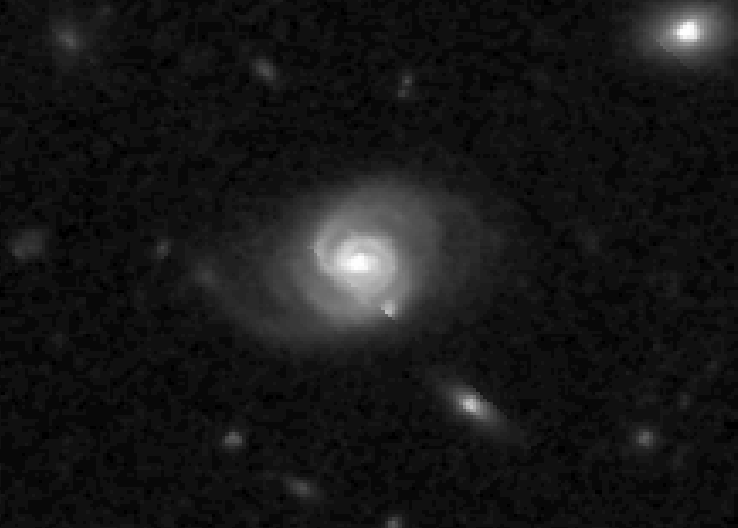}
        \includegraphics[width=8cm]{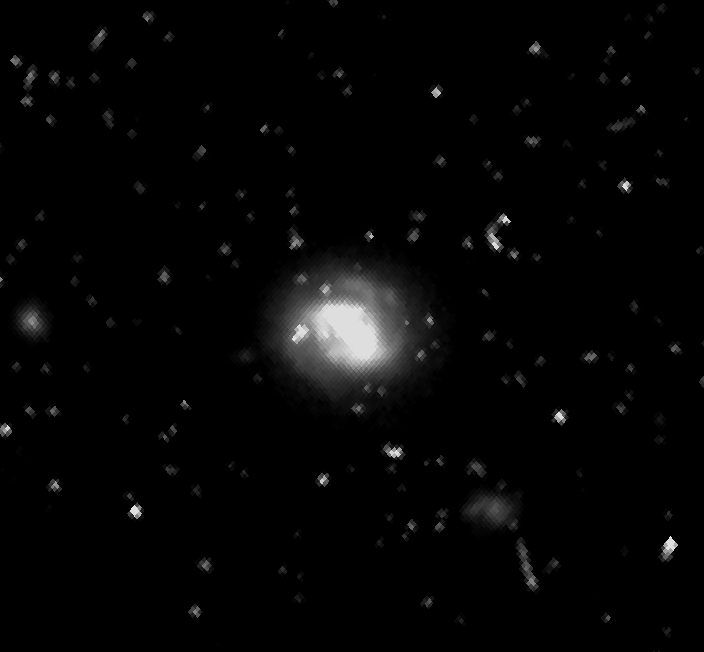}
        \includegraphics[width=8cm]{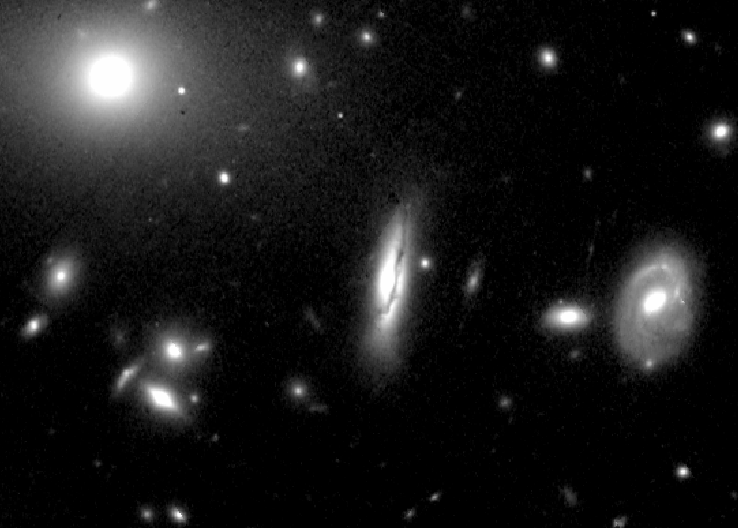}
        \caption{HST images of X-ray AGN hosts within 0.5$R_{500}$ radius. Top and middle: Two broad-line AGNs found in Abell 1689 and Abell 586 respectively. Their morphology is elliptical. Bottom: A narrow-line AGN detected near the cluster centre of Abell 1763 (The BCG is located at the top left corner of the image), which is classified as a dusty S0.}
        \label{fig:morphology}
\end{figure} 

\begin{figure}
    \centering
        \includegraphics[scale=0.26]{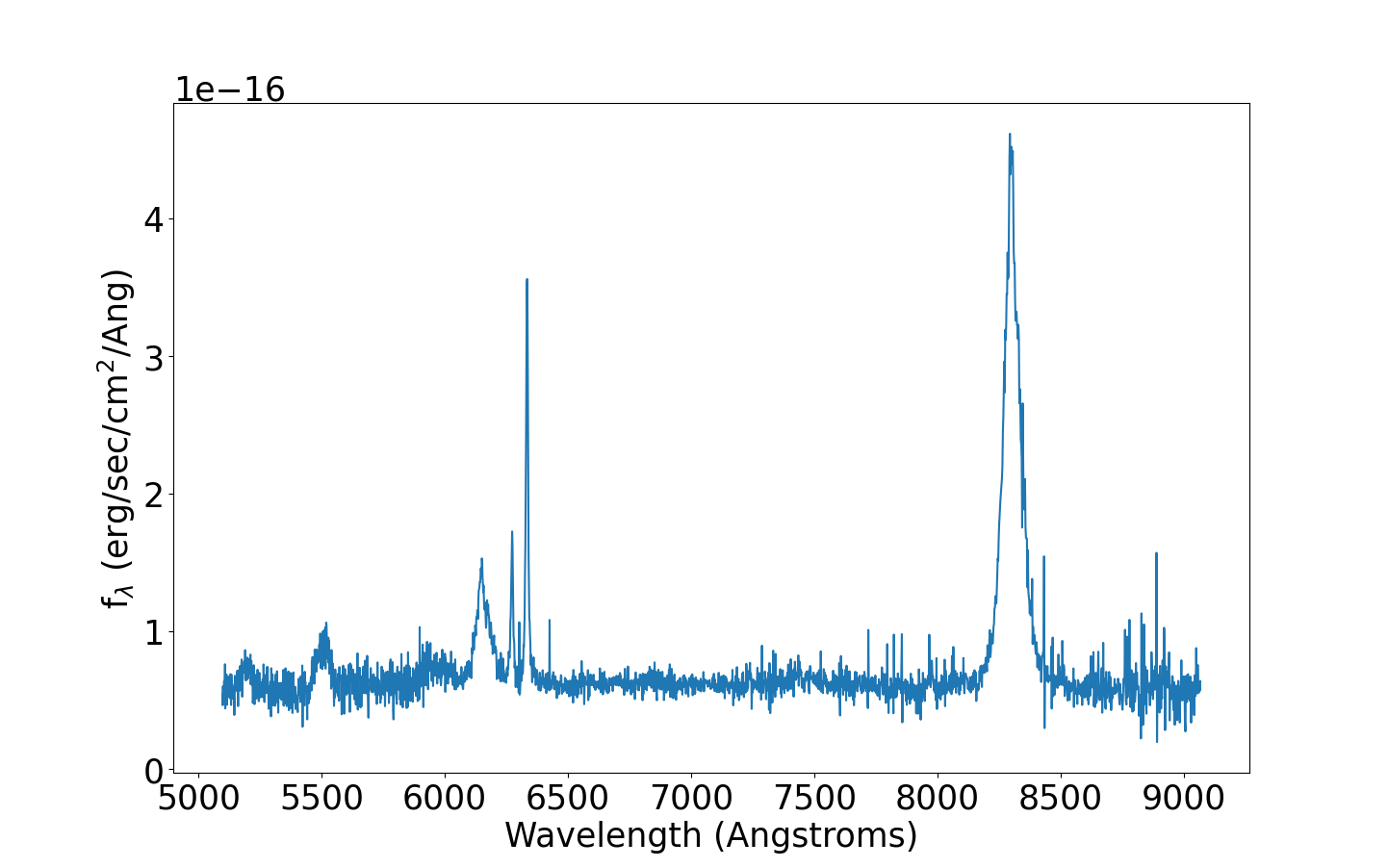}
        \caption{Broad-line optical spectrum of the host galaxy of the X-ray-detected AGN, Abell1835\_4.}
        \label{fig:broadline}
\end{figure}

\begin{figure}
\centering
        \includegraphics[scale=0.35]{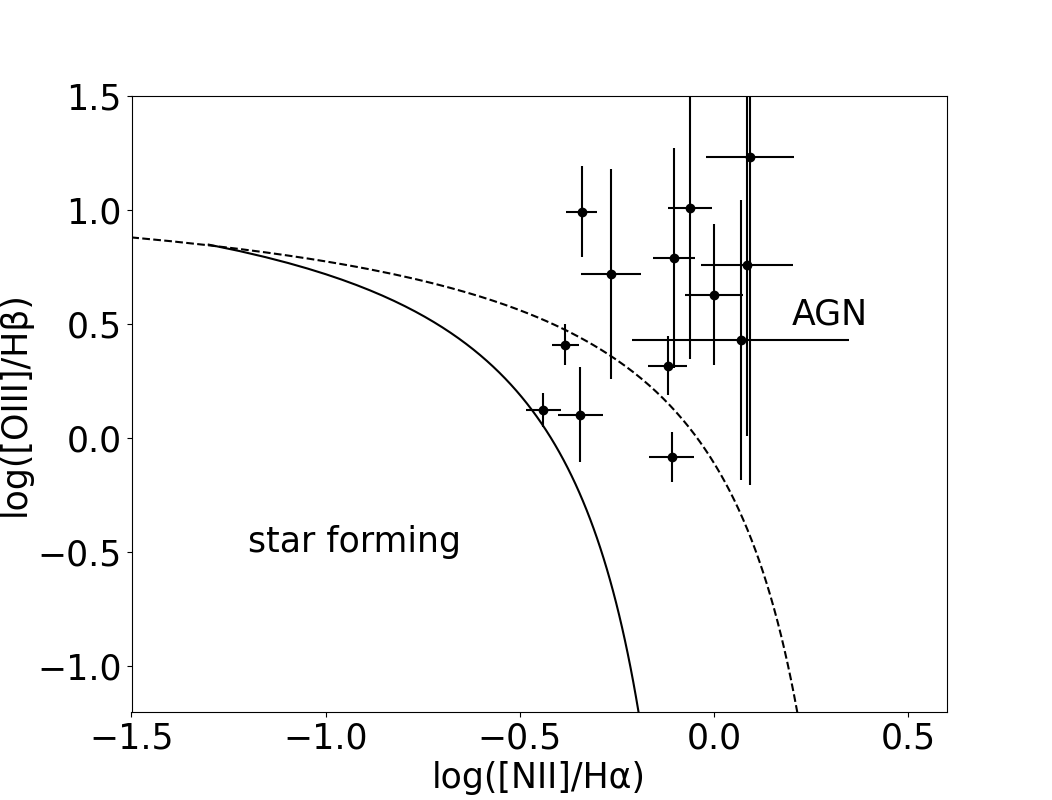}
        \caption{BPT diagram of the 13 sources with prominent narrow emission lines in their optical spectra (FWHM(Balmer)$<$500 km/sec). The continuous curve denotes the star-forming and AGN division of \citep{Kauffmann2003}, while the dashed curve the respective source separation of \citep{Kewley2001}. Based on this scheme, all the sources fall in the AGN part (above the continuous curve), while four of them may also include contribution from star formation (between the curves).}
        \label{fig:bpt}
\end{figure} 

\begin{figure}
\centering
        \includegraphics[width=7.5cm]{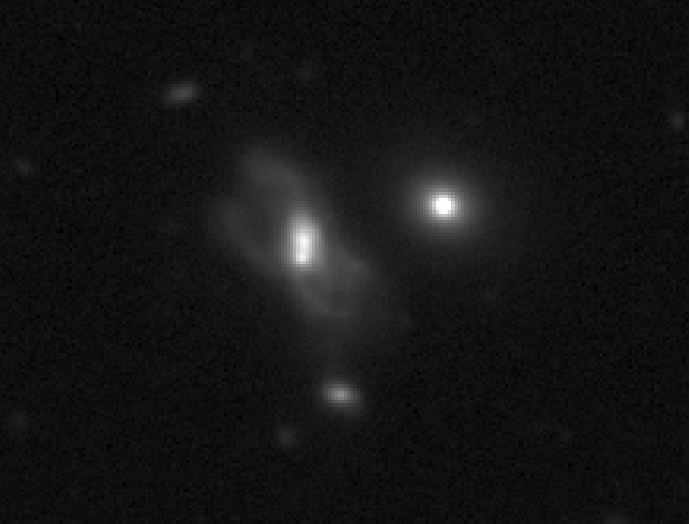}
        \includegraphics[width=7.5cm]{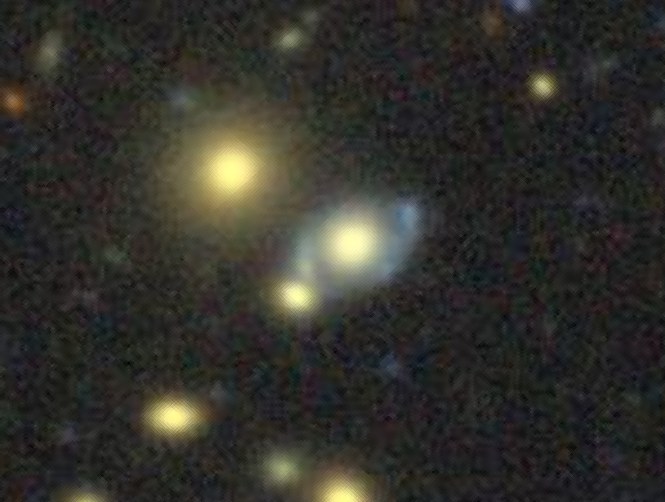}
        \includegraphics[width=7.5cm]{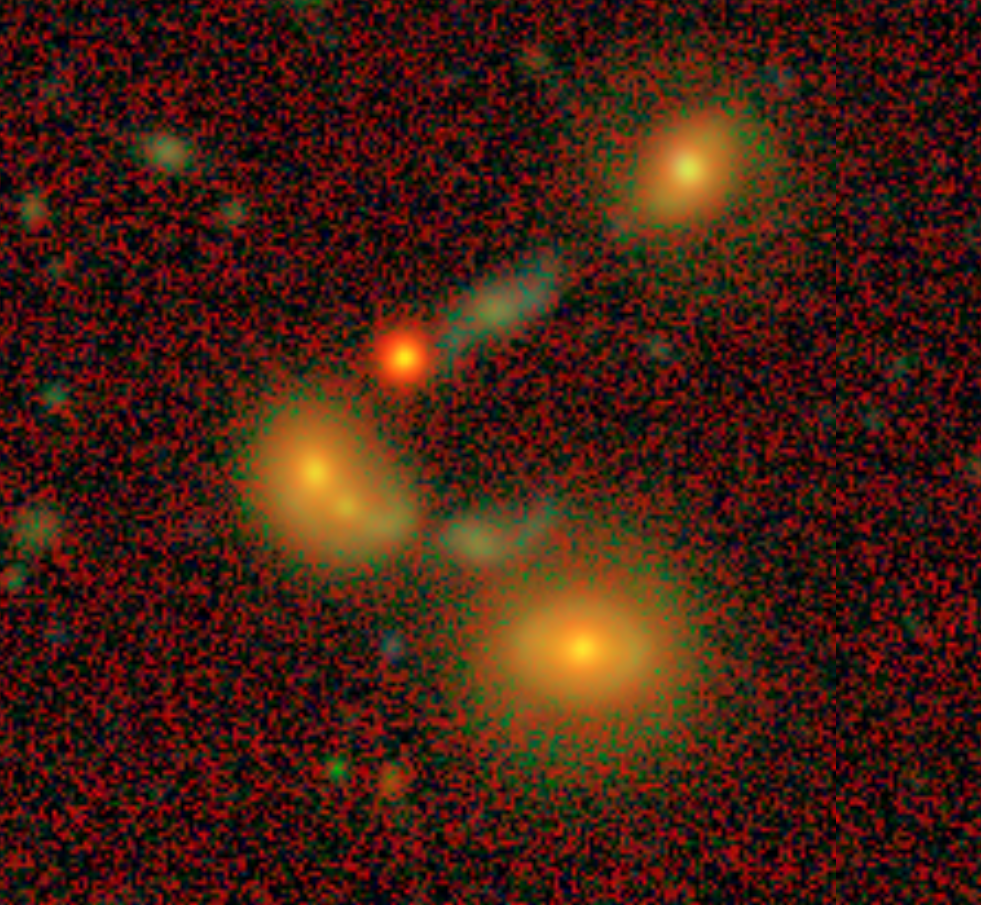}
        \caption{Images of irregular and interacting X-ray AGN host galaxies. Top: i-band HSC image of a disturbed galaxy in the outskirts of Abell 1835. It is probably in an advanced merging state and two nuclei are visible. Middle: DES-DR2 colour image of a possibly interacting galaxy in the outskirts of Abell 209. A circum-galactic ring of star formation is clearly visible. Bottom: HSC colour image of a possibly interacting group of galaxies in the outskirts of Abell 1835. The X-ray AGN host is the galaxy at the bottom of the image.}
        \label{fig:disturbed}
\end{figure} 

\subsection{Optical spectra of X-ray-selected AGNs} \label{sec:optical}

The majority of the optical counterparts of our X-ray-selected AGNs have prominent emission lines in their spectra. In more detail, the optical spectra of 17 out of the total 30 X-ray AGNs have strong emission lines, while another nine present weak emission lines (ELG), especially in the H$\alpha$ region; however, they are mostly dominated by the stellar continuum of an early type galaxy. The remaining four X-ray AGNs used in the current study have an absorption-line galaxy (ALG) spectrum typical of elliptical galaxies.    

Furthermore, we divided all optical AGNs with strong emission lines in broad-line (type-1 AGNs) and narrow-line (type-2 AGNs) sources based on the width of the Balmer lines. All spectra in the redshift range of this study include both the H$\beta$ and the H$\alpha$ region. In Fig.~\ref{fig:broadline} we present the broad-line spectrun of A1835\_4. We chose to consider all sources with a FWHM of the Balmer lines less than $500$ km/s and no evidence of any broadening with respect to the forbidden lines ([OIII], [NII]) as narrow-line
AGNs. We found 13 narrow-line sources, while the remaining four are classified as broad-line AGNs. To determine whether the emission-line spectra of the 13 narrow-line sources are consistent with an optical type-2 AGN classification, we plotted these sources on the BPT diagnostic diagram \citep{Baldwin1981}. The results are presented in Fig.~\ref{fig:bpt}. Based on this plot, the emission lines of all sources are produced, at least partially, by an AGN. Four out of the 13 narrow-line sources would be classified as composite sources, meaning they are partially ionized by star formation. The classification of all sources is included in Table~\ref{AGNlist}.

The ratio of type-2 to type-1 AGNs found in the current study is not dissimilar to the respective field ratio in the Local Universe \citep[e.g.][]{Maia2003}, or in high-redshift ($z\sim1$) clusters \citep{Mo18}. However, it is not in agreement with
recent results from local clusters in WINGS and Omega-WINGS surveys, where they found a very low optical type-2 to type-1 fraction ($\sim10$) compared to the field \citep{Marziani2023}. Nevertheless, the selection of the above mentioned samples are markedly different and any comparison should be taken with caution.
   
Interestingly, the majority of the broad-line AGNs (three out of four) are found within the central $0.5\times R_{500}$ radius. In addition, the spectra of all nine X-ray AGNs that are found within this central annulus are dominated by AGN emission lines, while moving towards the outskirts we find the ones dominated by the galaxy. The same is valid for the X-ray luminosity, as the most X-ray luminous sources are found within the first annulus. This is not unexpected considering that, because of the diffuse X-ray emission at the location of the cluster centre, the detection of low X-ray luminosity AGNs may be hampered even with the high resolution of {\it Chandra}, as also noted in \citet{Haines2012}. Most of the sources in the innermost annulus have a high LoS velocity relative to the cluster redshift, which indicates that they are an infalling population that entered the cluster potential within the last 700 Myr -- and not virialized cluster members accreted at earlier times.   

\section{Summary and discussion}
\label{sec:disc}

In this study, our goal has been to investigate the role of specific cluster properties in the triggering or suppression of AGN activity. To this end, we used a sample of 19 well defined massive clusters from the LoCuSS survey, observed by both {\it XMM-Newton} and Chandra observatories, and covered by extensive spectroscopic observations by the Hectospec spectrograph mounted on the MMT telescope. The advantage of this sample (although it is not large) is that it has undergone a thorough study; therefore, the basic cluster properties, such as the characteristic $R_{500}$ radius, dynamical state, and mass are very well defined. This is probably critical for this kind of studies \citep[see discussion in][]{Koulouridis2019}. Equally important is the highly complete spectroscopic follow-up of cluster galaxies below a magnitude limit of $M_K^*+1.5$. This allowed us to accurately define the fraction of X-ray AGNs in cluster galaxies, but also study their optical counterparts. Furthermore, the detection of all the infalling X-ray groups in these clusters enabled the association of their position with the X-ray AGNs and whether they may be triggered in these groups before entering the cluster potential.

Another advantage of this sample is that by combining the deep {\it XMM-Newton} and the Chandra observations in these fields we were able to detect AGNs out to $2\times R_{500}$ radius for almost all our clusters using the 4XMM DR10 catalogue, and at the same time not be severely hampered by the diffuse X-ray emission in cluster centres by using the Chandra archive of point-like sources.  

Our main results can be summarized as follows: (i) The X-ray AGN fraction is found consistent with the field in the outskirts of our clusters and it is decreasing towards their centres. This behaviour is independent of cluster mass within our mass range of $2\times10^{14} M_\sun<M_{500,x}<2\times10^{15} M_\sun$. (ii) The existence of infalling groups is not related to the triggering of AGN activity. On the contrary, X-ray AGNs are rarely associated to any of the detected X-ray clusters. (iii) CC clusters present a significant excess of X-ray AGNs in their outskirts, compared to non-CC clusters but also to the field. (iv) The majority of the optical spectra of the detected X-ray AGNs present emission lines. Out of the total 30 detected cluster AGNs, we found four broad-line sources. In the following sections, we discuss these findings. 

Previous studies have shown that not only the fraction of X-ray AGNs in massive clusters is significantly lower than in the field \citep[e.g.][]{Koulouridis2010, Haines2012}, but also that it is decreasing as we move from the outskirts towards the cluster centre \citep[e.g.][]{Haines2012,Ehlert2013,Ehlert2014,Koulouridis2018b}. Ram pressure stripping is considered to be the main driver of this behaviour. In more detail, the external pressure, $P$, which is able to strip the galaxy of its gas when it overcomes the gravitational forces that keep the gas anchored to the stellar disc of the galaxy is given by the relation:
 \begin{equation*}
     P=\rho_{ICM}V^2,
 \end{equation*}
where $\rho_{ICM}$ is the density of the ICM and $V$ is the galaxy velocity relatively to the ICM. The velocity of the galaxy within the cluster potential depends on the mass and it reaches its maximum value after the first infall near the pericenter. On the other hand, the density depends on the cluster-centric distance, $\rho_{ICM}\propto r^{-3\beta}$, but not on the cluster mass according to the self-similar scenario. However, observations showed that, while this is probably valid outside the $R_{500}$ radius, the density is decreasing with decreasing mass closer to the cluster centre. 
In a nutshell, ram pressure is higher in more massive structures and closer to the cluster centre. Furthermore, it affects more strongly galaxies closer to the cluster centre after their first infall. Since the AGN duty cycle is far shorter than the average crossing time, the decrease in AGNs toward the cluster centre probably indicates the lack of gas reservoir capable of inducing new cycles of activity. This picture is consistent with what we see in observations.

Nevertheless, the cluster centre is not devoid of X-ray-detected AGNs. This is not unexpected since RPS also depends on galaxy properties. In more detail, to obtain an effective stripping the drag force on the galaxy gas should overcome the forces that keep the gas bound to the galaxy, thus:
\begin{equation*}
    \rho_{ICM} V^2 > 2\pi G \Sigma_{star} \Sigma_{gas},
\end{equation*}
where $\Sigma_{star}$ is the stellar surface density and $\Sigma_{gas}$ the gas surface density. As shown in Figure 3 of \citet{Boselli2022}, while a massive cluster is able to completely strip dwarf galaxies ($M_{star}<8.3 M_\sun$) even before they cross the virial radius, more massive galaxies are able to retain at least some of their gas reservoir. Simulations showed that even dwarf galaxies can retain gas in the nucleus \citep{Steyrleithner2020}. Theory and simulations predict that this gas can lose angular momentum as a result of the stripping process and fall toward the SMBH, thus triggering the nuclear activity \citep{Schulz2001,Tonnesen2009,Ramos-Martinez2018}. This is supported by observations of cluster galaxies in the GASP survey \citep{Poggianti2017}, which indicated that the incidence of optical AGNs in "jellyfish galaxies" (for a precise definition see \citet{Ebeling2014}) is significantly higher than in non-cluster galaxies \citep{Peluso2022}. This is observed only in high-mass galaxies, while jellyfish galaxies probably comprise a small fraction of the total population of RPS galaxies. Furthermore, their characteristics may also be due to tidal forces in the gravitational potential of massive clusters ($M>10^{15}\;\rm M_\sun$) \citep{Boselli2022}, which are stronger for the infalling population. Therefore, we argue that, while RPS near the cluster centre may trigger the nuclear activity particularly in massive infalling galaxies, the gas loss in the infalling galaxy population is capable of decreasing the total number of AGNs.   
  
Although X-ray AGNs are rare in galaxies near the cluster core, most of them present strong emission lines in their optical spectra, while three out of the four broad-line AGNs of our sample were found within this region, lacking therefore evidence of nuclear absorption. In contrast, in the outskirts we find only one broad-line AGN, while almost half of the sample does not even present strong emission line spectra, which is instead dominated by the stellar continuum. This could indicate a different AGN-triggering mechanism or a difference in the accretion rate, $\dot m$, between the two regions. We argue that close to the cluster core, AGN activity may be triggered either by the influence of strong RPS, as suggested by studies of "jellyfish" galaxies \citep{Poggianti2017, Peluso2022}, or by tidal shocks, as galaxies pass through cluster pericentre. In the outskirts two different mechanism may play an important role. In more detail, \citet{Haines2012} argued that there are two distinct populations of X-ray AGNs in the outskirts, the newly infalling galaxies and the ones that have already passed the pericenter and they approach the apocenter of their orbit. The latter, the back-splash population, are dim in the infrared as a result of the gas loss that they have undergone during their first fast passage through the cluster. Their velocities of these galaxies should also be lower than of the infalling ones probably allowing for more interaction and merging with other galaxies in the outskirts. This could lead to the triggering of their AGN activity, albeit with a low $\dot m$, leading to an obscured nuclear activity or to an AGN that intrinsically lacks the broad-line region, the ones called true or naked type-2 \citep[e.g.][]{ElitzurTrump2014, Koulouridis2014, Koulouridis16a}. Nuclear activity may also be triggered in the outskirts by interaction of the host galaxy with the cluster itself, either as they pass through virial shocks \citep[e.g.][]{Keshet18} or via compression of gas onto the nucleus in the early stages of RPS. 

The back-splash population would also explain the high fraction of X-ray AGNs found in the current sample in the outskirts of CC clusters, as compared to non-CC clusters and to the field. In these more relaxed clusters probably the galaxies have more time to reach the apocentre and spend more time there without the disrupting events of cluster merging, as shock waves in the ICM generated by cluster mergers may enhance the RPS \citep[e.g.][]{Vijayaraghavan13,Jaffe16}. This is in agreement with \citet{Ruderman2005}, who discovered a mild excess of X-ray sources between 1.5 and 3 Mpc in 24 relaxed massive clusters spanning the redshift range $z$ = 0.3-0.7. Similarly, \citet{Stroe2020} found that also the population of H$\alpha$-selected AGNs in 14 galaxy clusters peaks at the same region ($\sim1.5-3$ Mpc) of their relaxed subsample. Interestingly, in the outskirts of the most massive CC cluster of our sample, Abell 1835, we found seven X-ray-detected AGNs, while none is located within the central $R_{500}$ radius. In sharp contrast, in Abell 1758, also a very massive cluster, albeit without a CC and in a state of merging with another massive cluster, we found only one X-ray-detected AGNs very close to the cluster centre. 

The morphology of AGN hosts may hint on the possible mechanism that is responsible for the AGN activity in cluster members. While we lack homogeneous imaging for the whole sample, many of our X-ray-detected AGNs are covered by HST and HSC (Hyper-Suprime-Cam), allowing for detailed morphological classification. The rest are covered by DES, DECaLS, and PanSTARRS. The optical morphology classification is presented in Table~\ref{AGNlist}. An interesting result is that two out of the three broad-line AGNs that are located very close to the cluster core are hosted in spiral galaxies (HST imaging), indicating that they are found in infalling galaxies during their first approach to the pericentre of their orbit. The third one is unresolved due to its proximity to a very bright star. The only broad-line AGN in the cluster outskirts is hosted by what is most likely an interacting E/S0 galaxy (HSC imaging, see Fig.~\ref{fig:morphology}).   

Also, we have found several AGNs hosted by galaxies with highly disturbed or irregular morphology, which indicates recent interactions or merging. Most of these cases are found in the outskirts and some examples are shown in Fig.~\ref{fig:disturbed}. A higher merging rate was also found in a recent study of AGNs in XXL survey clusters compared to field AGNs and cluster galaxies with no AGN activity (E. Drigga et al. in prep.). It is also evident that emission-line AGNs are hosted in spiral or disturbed and irregular galaxies, while undisturbed early-type galaxies host X-ray AGNs that are either lacking  or with very weak emission lines. Although our AGN sample is small, these results indicate that the AGN activity close to the cluster centre may be triggered by RPS in infalling galaxies; in the outskirts, it is the outcome of merging and interactions between galaxies with lower velocities, less gas, and older stellar populations. Most of the disturbed and irregular galaxies host an AGN with a narrow-line spectrum, which hints towards nuclear obscuration caused by interactions. This is also supported by their X-ray spectrum, which (in many cases) is very bright in the hard [2.0-10] keV band, whereas it is far weaker in the soft X-ray [0.5-2.0] keV band, and frequently lower than the selection threshold ($L_x=10^{42}$ erg\;sec$^{-1}$). In addition all disturbed and merging hosts, as classified by visual inspection, are the brightest infrared sources in the four WISE bands.    

Finally, we find no correlation between the positions of the X-ray-detected infalling groups and the X-ray AGNs. However, simulations recently showed that infalling galaxy groups are disrupted before the pericenter and only galaxies closest to the group center,<$0.7R_{200}$, may remain bound to the group \citep{Haggar2023}. They also argue that these still bound galaxies are slow-moving and thus they can be tidally disrupted or in a merging state more easily. In addition, \citet{Vijayaraghavan13} showed that the merging rate of the infalling group galaxies steadily increases until the first pericentric passage. To test this result, we used our data to investigate the morphology of the X-ray AGN hosts that are located within $R_{200}$ in any infalling group. We found three X-ray AGNs that satisfy the above criterion and two of those are indeed likely to be interacting with a nearby neighbour (see middle and bottom panels of Fig.~\ref{fig:disturbed}). Nevertheless, most of our X-ray-detected AGNs and most of the disturbed or merging cases are not found anywhere near the X-ray-detected infalling groups.   

\section{Conclusions}
\label{sec:conc}

The  conclusions of our analysis of X-ray-detected AGNs in 19 massive galaxies are given below. 
\begin{itemize}

\item The X-ray AGN fraction in cluster galaxies is higher in the outskirts of clusters than in the centres, which is in agreement with previous results for massive clusters. This does not depend on cluster mass within the mass range of the current study, $M_{500,x}>2\times10^{14}\,M_\sun$. Furthermore, there is no evidence that the presence of X-ray-detected infalling groups plays any role in the prevalence of X-ray AGNs in clusters. 
        \newline

\item An excess of X-ray AGNs is found in the outskirts of relaxed clusters, compared both to non-relaxed clusters and to the field. The disturbed morphology of several galaxies in the outskirts indicates that it can be due to an enhanced galaxy merging frequency.
\newline
    
\item According to previous results, the fraction of broad- to narrow-line AGNs in clusters is roughly consistent with the corresponding fraction in the field. However, broad-line AGNs may be preferably located closer to cluster centres.
   \newline
 
\item The difference in the fraction of X-ray AGNs between outskirts and cluster centres is probably due to the different prevailing physical mechanism that affects their triggering. Ram pressure stripping is likely to be strongly affecting galaxies closer to the cluster centre, while in the outskirts of relaxed clusters, frequent galaxy mergers are dominant. 

\end{itemize}

Future research should aim to maximise AGN sample sizes within meticulously characterised cluster datasets. Our future plans involve the scientific exploration of extensive X-ray selected cluster samples, including but not limited to X-CLASS \citep{Koulouridis2021} and XCS \citep{Mehrtens2012, Giles2022}.

\begin{acknowledgements}
The authors extend their gratitude to the anonymous referee for the attentive review and valuable feedback. The authors express their gratitude to Professor Eiichi Egami for granting them access to the ACReS database. EK acknowledges support under the grant 5089 financed by IAASARS/NOA. ED acknowledges financial support by the European Union’s Horizon 2020 programme “XMM2ATHENA” under grant agreement No 101004168. This research has made use of "Aladin sky atlas" developed at CDS, Strasbourg Observatory, France. This research has made use of ESASky \citep{Baines2017,Giordano2018}, developed by the ESAC Science Data Centre (ESDC) team and maintained alongside other ESA science mission's archives at ESA's European Space Astronomy Centre (ESAC, Madrid, Spain). This research made use of Astropy, a community-developed core Python package for Astronomy (http://www.astropy.org, Astropy Collaboration 2018). This publication made use of SAOImageDS9 \citep{Joye2003}. This publication made use of TOPCAT \citep{taylor2005} for table manipulations. The plots in this publication were produced using Matplotlib, a Python library for publication quality graphics \citep{Hunter2007}. Based on observations
obtained with {\it XMM-Newton}, an ESA science mission with instruments and contributions directly funded by ESA member states and NASA.
\end{acknowledgements}

\bibliographystyle{aa}
\bibliography{48212corr}
%
%
\begin{appendix}
\section{List of X-ray detected AGN}
In table ~\ref{AGNlist} we present the list of 30 X-ray detected AGNs in our cluster sample.

\begin{sidewaystable*}
\captionsetup{width=.9\textwidth}
  \centering
  \caption{List of X-ray detected AGNs. \\(1) Cluster name, (2) right ascension (X-ray position), (3) declination (X-ray position), (4) right ascension (optical position), (5) declination (optical position), (6) X-ray luminosity, (7) redshift, (8) morphological classification by visual inspection, (9) classification of the optical spectrum, as described in Sect. \ref{sec:optical}}.
  \label{AGNlist}
  \begin{tabular}{lrrrrcccc}
  \hline
  \noalign{\smallskip}
  Cluster& RA (X--rays) & \multicolumn{1}{c}{Dec (X--rays)} & \multicolumn{1}{c}{RA (opt.)} & \multicolumn{1}{c}{Dec (opt.)} & \multicolumn{1}{c}{$L_X$} & \multicolumn{1}{c}{$z$} & \multicolumn{1}{c}{Morphology} & \multicolumn{1}{c}{Optical spectrum} \\
  name & \multicolumn{1}{c}{(J2000)}  & \multicolumn{1}{c}{(J2000)}  &  \multicolumn{1}{c}{(J2000)}  & \multicolumn{1}{c}{(J2000)} & \multicolumn{1}{c}{(erg sec$^{-1}$)}  &  & (imaging survey)\\
   (1)& \multicolumn{1}{c}{(2)}&\multicolumn{1}{c}{(3)}&\multicolumn{1}{c}{(4)}&\multicolumn{1}{c}{(5}&\multicolumn{1}{c}{(6)}&\multicolumn{1}{c}{(7)}&\multicolumn{1}{c}{(8)}&\multicolumn{1}{c}{(9)}\\
  \hline
  \noalign{\smallskip}
  Abell 586     &       113.0712        &       31.6140 &       113.0710        &         31.6141 &       7.8$\times10^{42}$      &       0.174   &       Spiral (HST)  & Broad-line \\
Abell 963       &       154.2529        &       39.0757 &       154.2530        &         39.0758 &       3.3$\times10^{43}$      &       0.210   &       Unresolved (PanSTARRS) & Broad-line \\
   Abell 1689    & 197.8984 &   -1.3368 & 197.8984      & -1.3367       &         1.4$\times10^{42}$ & 0.200      & Spiral (HST) & Broad-line \\
   Abell 1758   & 203.1842 &    50.5188 & 203.1842 &    50.5188 &       2.4$\times10^{42}$ & 0.287   & unresolved (HST) & Narrow-line \\
   Abell 1763   & 203.8266      &40.9971        &203.8266&      40.9972&        4.4$\times10^{42}$ &0.240  & S0 (HST)  & Narrow-line \\
   Abell 2219   & 250.0242      &46.7281        &250.0244       & 46.7282 &       1.5$\times10^{42}$ & 0.229      &       unresolved (PanSTARRS) & Narrow-line \\
   Abell 2390   & 328.4082      & 17.7298&      328.4083        & 17.7299       &       1.6$\times10^{42}$ &  0.241         &      S0 (HST) & Narrow-line \\
   Abell 2390   & 328.4398      & 17.6966       & 328.4399      & 17.6966       & 3.1$\times10^{42}$ & 0.214      & Spiral/Irregular (DECaLS) & Narrow-line \\
   Abell 611    & 120.2377      & 36.1240       & 120.2375      & 36.1237       & 1.4$\times10^{42}$ &    0.283& E/S0 (PanSTARRS) & Narrow-line/no H$\beta$ region\\
    Abell 611   & 120.2593      & 36.0459       & 120.2593      & 36.0460       &         1.9$\times10^{42}$ & 0.276      & S0 (HST)  & Narrow-line \\
   \hline
   $R_{500}$\\
   \hline
   ZwCl0104.4+0048      &       16.6205 &       1.0933  &       16.6200 &         1.0924  &       2.9$\times10^{42}$      &       0.261   & E/S0 (HSC) &       ALG \\
   Abell 209    &       22.8890 &       -13.6198        &       22.8893 &         -13.6197 &      4.8$\times10^{42}$      &       0.203   &       E/S0 (DES) & ELG \\  
   Abell 209    &       22.8904 &       -13.5398 &      22.8908 &       -13.5398        &         5.1$\times10^{42}$      &       0.218   &       Spiral/merger (DES) & Narrow-line \\
   Abell 209    &       23.0676 &       -13.5937        &       23.0681 &         -13.5939 &              4.2$\times10^{42}$      &       0.203   &         merger? (DES) & ELG \\
   Abell 383    &       42.0182 &       -3.3740 &       42.0179 &       -3.3737 &                 9.3$\times10^{42}$      &       0.186   &       Irregular (DES) & ELG     \\
   Abell 383    &       42.1026 &       -3.5286 &       42.1028 &       -3.5292 &         3.0$\times10^{42}$      &       0.186   &       E/S0 (DES) & ALG \\
   Abell 963    &       154.3848 &              39.0204 &       154.3848 &       39.0206 &               3.4$\times10^{43}$      &       0.202   &         Unresolved (PanSTARRS) & Narrow-line \\
   Abell 1763   & 203.9821      & 41.0240       & 203.9812      & 41.0247       & 5.2$\times10^{42}$      & 0.235 & Unresolved (PanSTARRS) & ELG \\
   Abell 1835   & 210.2462      & 3.0132        & 210.2462      & 3.0132        & 3.7$\times10^{42}$ &    0.265   &       E/S0 (HSC) & ELG        \\
   Abell 1835   & 210.2134 & 2.9847     & 210.2138      & 2.9849        & 6.0$\times10^{42}$      & 0.256 & E/S0 (HSC) & Narrow-line \\
   Abell 1835   & 210.3171      & 2.7533 & 210.3173     & 2.7537        &         9.8$\times10^{42}$      & 0.245 & E/S0 (HSC) & ELG \\   
Abell 1835      & 210.3653      & 2.9350 &  210.3654    & 2.9350        &  1.4$\times10^{44}$             & 0.265 & Irregular/interacting? (HSC)  & Broad-line \\
Abell 1835      & 210.3686      & 2.8155        & 210.3684      & 2.8157        & 3.8$\times10^{42}$      & 0.245 & S0/interacting (HSC) & ELG \\
Abell 1835      & 210.4146      & 2.9559 &      210.4145        & 2.9560        &         9.5$\times10^{42}$      & 0.250 & E/S0 (HSC) & ELG weak \\
Abell 1835      & 210.4377 & 2.8923     & 210.4376      & 2.8926        &         6.5$\times10^{43}$      & 0.249 & Irregular (HSC)  & Narrow-line \\
Abell 2219      & 250.1797      & 46.6102       & 250.1803      & 46.6106 &       2.8$\times10^{42}$      &       0.232   & E/S0 (PanSTARRS) & ELG \\
RXJ2129.6+0005  &       322.3569 &      0.2020  &       322.3573 &      0.20191 &               8.2$\times10^{42}$      &       0.241 &         E/S0 (DES) & ALG \\
Abell 2390      & 328.5686      & 17.7388       & 328.5685      & 17.7389       & 2.3$\times10^{42}$      & 0.232 &       Unresolved (DECaLS)  & Narrow-line \\
Abell 2390      & 328.5138      & 17.6318       & 328.5136      & 17.6316       & 1.6$\times10^{43}$      & 0.236 &       Unresolved (DECaLS)  & Narrow-line \\
Abell 2390      & 328.3709      & 17.5161       & 328.3718      & 17.5165       & 2.4$\times10^{42}$      &       0.225   &       E/S0 (DECaLS) & E \\
\hline
\end{tabular}
\end{sidewaystable*}
\end{appendix}

\end{document}